\documentclass[aps,prl,superscriptaddress]{revtex4}

\usepackage{amsmath}
\usepackage{graphicx} 

\begin{document}

\title{Spatially resolved quantum nano-optics of single photons using an electron microscope }

\author{L. H. G. Tizei}

\affiliation{Laboratoire de Physique des Solides, Universit\'e Paris-Sud, CNRS-UMR 8502, Orsay 91405, France}

\author{M. Kociak}
\email{mathieu.kociak@u-psud.fr}
\affiliation{Laboratoire de Physique des Solides, Universit\'e Paris-Sud, CNRS-UMR 8502, Orsay 91405, France}

\date{\today}

\begin{abstract}
We report on the experimental demonstration of single photon state generation and characterization in an electron microscope. In this aim we have used low intensity relativistic (energy between 60kV and 100 keV) electrons beams focused in a ca 1 nm probe to excite diamond nanoparticles. This triggered individual neutral Nitrogen-vacancies  (NV$^{\text{0}}$) centers to emit photon which could be gathered and sent to a Hanbury Brown Twiss intensity interferometer. The detection of a dip in the correlation function at small time delays clearly demonstrates antibunching and thus the creation of non-classical light states. Specifically, we have also demonstrated single photon state detection. We unveil the mechanism behind quantum states generation in an electron microscope, and show that it clearly makes cathodoluminescence the nanometer scale analog of photoluminescence rather than electroluminescence. By using an extremely small electron probe size and the ability to monitor its position with sub nanometer resolution, we also show the possibility of measuring the quantum character of the emitted beam with deep sub wavelength resolution.

\begin{description}
\item[PACS numbers] 42.50.Dv, 42.50.-p, 78.60.Hk, 07.78.+s

\end{description}

\end{abstract}

\maketitle


The study of single photon sources has attracted great attention \cite{Glauber1963,Loudon2010, Lamb1947, Walls1983, Grangier1986, Gruber1997, Mizouchi2012, Lounis2005, Kok2007, Bennet1984}. The interest in these emitters stems from the mandatory requirement to create optical states which are fundamentally different from classical ones for fundamental \cite{Lounis2005} or technologically important applications, such as quantum computing and quantum cryptography \cite{Kok2007, Bennet1984}.  Reliable single photon sources (SPS) have been demonstrated based, among others, on Nitrogen-Vacancy (NV) centers in diamond, which have been extensively studied using photoluminescence (PL) \cite{Gruber1997} and in a lesser extent using electroluminescence (EL) \cite{Mizouchi2012} techniques. 
 
\begin{figure}
\includegraphics[width=1\columnwidth]{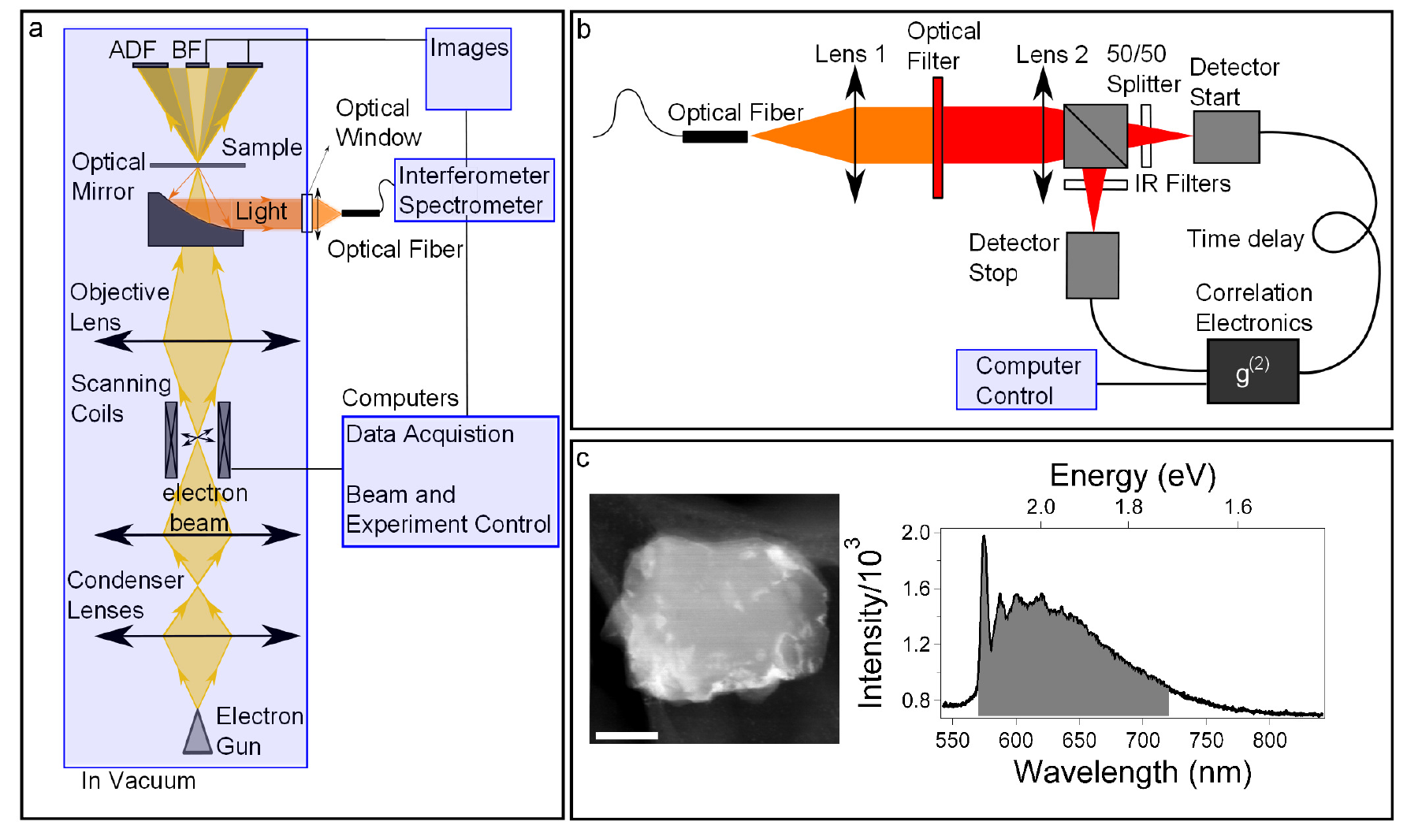}
\caption{\label{figure1}A light collection system was installed in a STEM microscope: a parabolic mirror collects the light emitted during electron irradiation, which is focused by a lens onto an optical fiber. This fiber is coupled to a spectrometer or to an HBT interferometer. b) HBT interferometer. c) ADF image of a diamond particle and a spectrum of light emitted from a particle containing NV$^{\text{0}}$ centers, detected between 570 nm and 720 nm (grey area). The scale bar in c) is 100 nm.}

\end{figure}

After excitation of an SPS, the probability of detecting two simultaneous photon emissions is zero, independently of the exciting probe statistics. This amazing quantum effect is called “photon antibunching”. Its observation unambiguously confirms the detection of quantum states of light. Single photon states can be evidenced by measuring the second order correlation function, $g^{(2)}(\tau)$. This function,

\begin{equation}
 g^{(2)}(\tau)= \frac{\langle I(t)I(t+\tau) \rangle}{\langle I(t)\rangle\langle I(t+\tau)\rangle}
\label{equation1}
\end{equation} 

provides information about intensity, $I(t)$, correlations of a given light field at different time delays, $\tau$. It can be measured using a Hanbury-Brown and Twiss (HBT) intensity interferometer (see below and Fig. \ref{figure1}). For classical light, $g^{(2)}(\tau) \geq 1$ for any $\tau$. However, for a single photon beam, $g^{(2)}(0) = 0$ \cite{Loudon2010, Gruber1997}. That is, given that a detection event has taken place, the probability of a second detection (for times shorter than the lifetime of the emitter) is lower than that for classical light. More generally, $g^{(2)}(0)$ scales as $(1-1/n)$, where $n$ is the number of photons in the state.

The requirement for the emission of one photon at a time implies that the emitting object will be ideally a two levels system in which saturation effects ensures that the system cannot be re-excited, unless a photon is previously emitted. Thus, generally, an SPS will have a limited size: an atom (e. g. Cs atoms in an optical cavity \cite{McKeever2004}), a point defect (e. g. NV in bulk diamond \cite{Gruber1997}) or a quantum dot (e. g. GaN in a AlN matrix \cite{Kako2006} or CdSe nanocrystals \cite{Spinicelli2009}), etc. Incidentally, conventional optical techniques, being diffraction limited \cite{Novotny2006}, will probe single objects only in highly dispersed samples. Sub-wavelength photon-based microscopy may suffer from other limitations: scanning near field optical microscopy signal decreases rapidly with spatial resolution \cite{Okamoto2006} and stimulated emission depletion microscopy cannot selectively excite multiple quantum emitters inside small objects \cite{Rittweger2009, Greffet2012}. However, advances in understanding of the physics of light at nanometric scales are clearly desirable. Hence, techniques assessing quantum optics at the scales relevant to many objects and at which their interactions take place are necessary.



In this letter, to circumvent the described difficulties, we have used a radically different approach to the pure optical means. We have have demonstrated the use of fast electrons (relativistic particles with energy set between 60 kV and 100 kV) focused in  a c.a. 1 nm-wide beam formed in a scanning transmission electron microscope (STEM) to excite neutral (NV$^{\text{0}}$) centers in diamond nanoparticles and prove that they can trigger SP emission. The de-excitation mechanism is evidenced to be equivalent to that of photoluminescence through the center’s lifetime. It  indicates that cathodoluminescence at the nanometer scale can be seen as a broadband analog of photoluminescence rather than electroluminescence. We also showed that the excitation position can be controlled to allow the deep sub wavelength characterization of the quantum character of the emitted states.

The experiments have been performed in a VG HB 501 STEM microscope with an in-house made scanning electronics. This microscope is equipped with a liquid Nitrogen cooled sample stage. Light (cathodoluminescence \cite{Yacobi1990}, CL) was collected with a carefully optimized high efficiency collection system \cite{Zagonel2011}, which was crucial to the success of the experiments. Light was coupled to a HBT interferometer \cite{Hanbury1956} or a spectrometer (Fig. \ref{figure1}a-b) using an optical multimode fibre (100 $\mu$m diameter core). Single photons have been detected by two Picoquant’s tau-SPADs (single photon avalanche diodes). Time delay histograms (which are proportional to the second order correlation function) have been acquired using the Time Harp correlation electronics, from Picoquant. The typical room background noise varied between 100 count/s and 500 count/s. Wavelength filtered images have been acquired by measuring a SPAD count signal. The typical acquisition time for each correlation curve was 300 s. Time delay histograms have been normalized to one for $\tau \gg 0$. This is justified by the shape of the curves, which do not show a bunching effect. Opposite to conventional PL, standard background subtraction has not been performed because it cannot be estimated from the collected data. Light intensity and statistics are not constant within a particle, rendering unjustifiable the usual \cite{Gruber1997, Beveratos2002} subtraction of a poissonian background without further information. At each position of a scan of the electron beam, two structural signals (annular dark field, ADF, roughly proportional to the projected mass, and Bright Field, BF), the light emission spectra (Fig. \ref{figure1}c) or the $g^{(2)}(\tau)$ can be acquired in parallel. This allows the localization, with nanometre accuracy and without any ambiguity, of the light emission property within the object of interest (For example, using this setup, without the HBT interferometer, we have recently investigated different color centers in diamond nanoparticles \cite{Tizei2012}). In other words, it gives access to the advantages of well established electron microscopy techniques and a quantum optic setup in a single experiment. 
To extract the antibunching dip depth histograms have been fitted to the following model \cite{Beveratos2002}:

\begin{equation}
g^{(2)}(\tau) =
\left\{
	\begin{array}{ll}
		1-g*e^{-(\tau-\tau_0)/\Gamma}  & \mbox{if } \tau \geq 0 \\
		1-g*e^{(\tau-\tau_0)/\Gamma}  & \mbox{if } \tau < 0
	\end{array}
\right.
\label{equation2}
\end{equation}

where $(1-g)$ is the depth of the antibunching dip, $\Gamma$ is the de-excitation lifetime of the center and $\tau_0$ is the position of the minimum value of the $g^{(2)}(\tau)$ function.

Two kinds of nanoparticles have been used: 1) a sample of large diamond particles (larger than 500 nm) crushed using a mortar (Aldrich); and 2) a sample with diamond nanoparticles in the 100 nm to 200 nm range (Microdiamant). Both samples have been diluted in deionised water and dispersed on holey-carbon copper grids.
Among those nanoparticles, for the presented experiments, SP candidates have been chosen based on emission intensity in the wavelength range of interest. After each measurement an ADF image was acquired to check that the nanoparticle had not drifted.



\begin{figure}
\includegraphics[width=1\columnwidth]{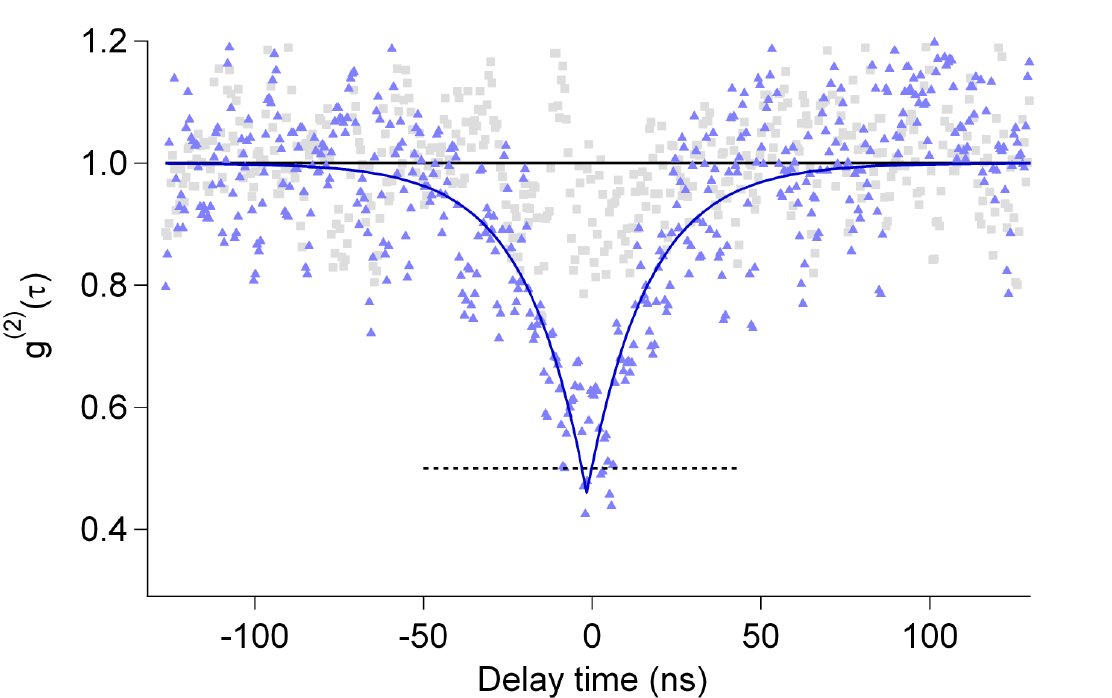}
\caption{\label{figure2} $g^{(2)}(\tau)$ curves measured in two different nanoparticles excited by fast electrons. The black curve was measured in a particle emitting 100 kcount/s and does not show light antibunching. The blue curve was measured in a particle emitting 30-40 kcount/s and clearly shows light antibunching. This is a clear signature of the quantum nature of the light beam. As $g^{(2)}(0) = (0.46 \pm 0.05)$ this also demonstrates the detection of a single photon emitter in the diamond lattice.}

\end{figure}

\begin{figure}
\includegraphics[width=1\columnwidth]{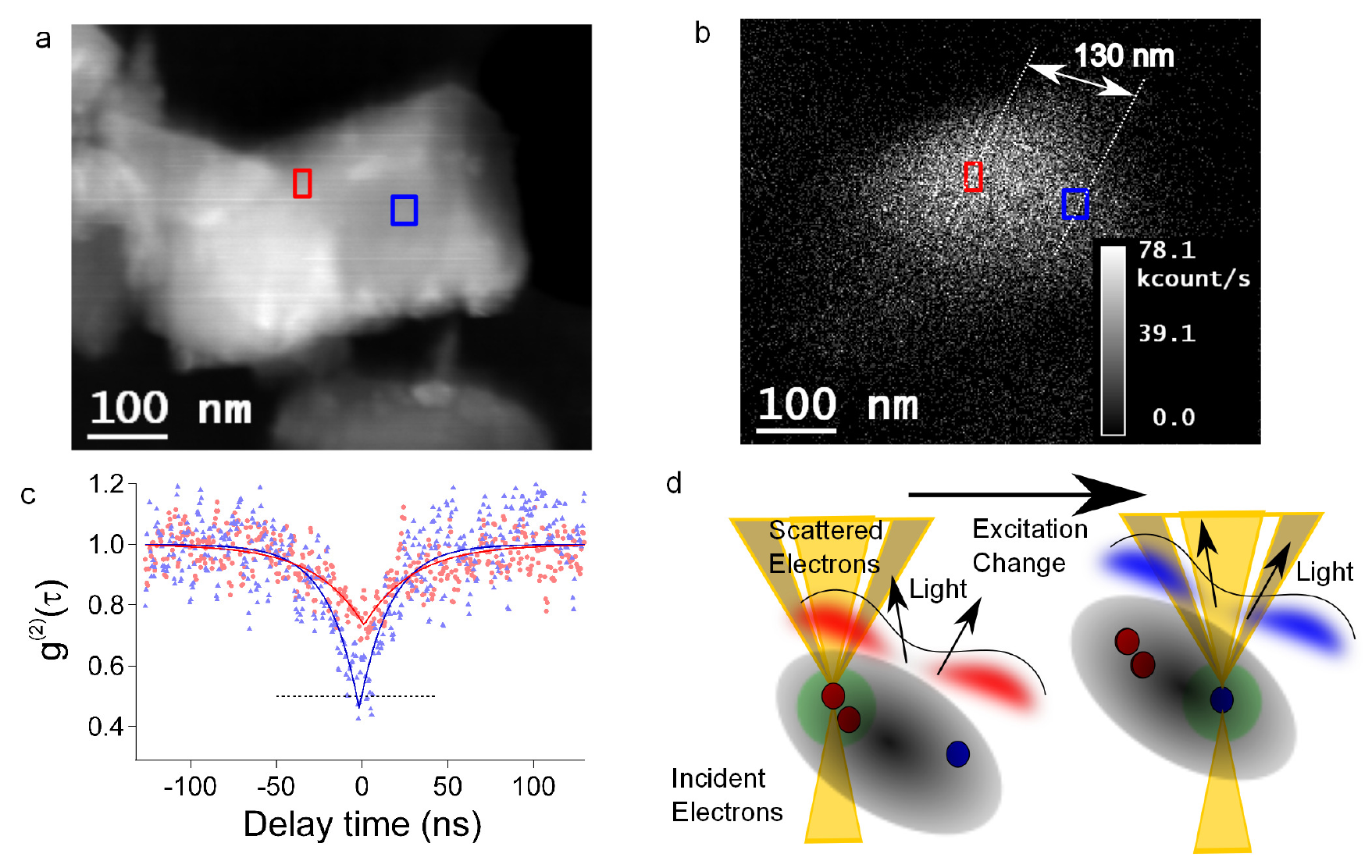}
\caption{\label{figure3} a) ADF and b) integrated light intensity images of a diamond nanoparticle acquired in parallel (exposure time 64 µs per pixel). Light emission is not homogeneous through the particle. c) $g^{(2)}(\tau)$ acquired with the electron beam scanning different regions (symbols represent data and lines fits), separated by 130 nm and marked by rectangles of the corresponding colors. The measured $g^{(2)}(0)$ values are $(0.46 \pm 0.05)$ and $(0.73 \pm 0.02)$. d) An electron beam excites a volume, defined by the electron probe and diffusion of carriers (light green circle). By shifting the probe one can excite light emission from another region, with possibly corresponding different $g^{(2})(\tau)$ value.}

\end{figure}

We could observe two different statistical properties for the investigated particles. $g^{(2)}(\tau)$ measured from bright diamond nanoparticles (100 kcount/second per detector or more) are flat ($g^{(2)}(\tau) = 1$) within our experimental time resolution (around 350 ps). Such behaviour indicates a classical source; see the black curve of Fig. \ref{figure2}. On the other hand, measurements on nanoparticles with weaker light emission (ca 30 kcount/s) show antibunching (blue curve in Fig. \ref{figure2}). This undoubtedly demonstrates the detection of non-classical light generated by fast electrons, which has not been reported so far to the best of our knowledge. The observation of classical curves in exactly the same experimental conditions, but for other nanoparticles, guarantees that the observed anti-bunching is not an effect of the electron beam statistics. The electron beam was maintained scanning a small fixed area (30 nm by 34 nm wide) on the nanoparticle (small blue rectangle on Fig. \ref{figure3}a). The difference between the two statistical behaviours can be explained as due to different concentrations of NV$^{\text{0}}$, or other centers, following the behaviour in PL and EL experiments, in which antibunching is seen only when few NV$^{\text{0}}$ are excited (and in the absence of other centers).
 
We measured $g^{(2)}(0) = (0.46 \pm 0.05)$ for the blue curve. $g^{(2)}(0) < 0.5$ is the demonstration that a beam of single photons has been detected.  The non-zero value can be explained by the presence of a background signal \cite{Beveratos2002}. In our experiment, the main origin of the background is the excitation of other centers due to charge carrier diffusion \cite{Tizei2012}. The deduced lifetime, $\Gamma = (18 \pm 4) ns$ is compatible with values for NV$^{\text{0}}$ centers in nanodiamond obtained by PL measurements and is a fundamental difference between the presented experiments and EL \cite{Mizouchi2012} for SP generation. The measured $g^{(2)}(\tau)$ unambiguously shows that we have detected single photons states emitted from a diamond nanoparticle excited by a nanometre-wide fast electron beam

Of course, the measurement of $g^{(2)}(0) < 1$ on an individual particle can also be achieved by optical techniques provided it is well separated from its neighbours. However, optical techniques do not give access to information about variations of the $g^{(2)}(\tau)$ within the same nanoparticle, or more generally at deep subwavelength spatial resolution, nor do they give parallel access to the nanometre resolved image. Here, the excitation by fast electrons focused into a small probe onto a thin specimen provides the breakthrough needed to measure these variations. The remarkable similarities between the present experiments and PL, in contrast to EL \cite{Mizouchi2012}, are explained by the fact that fast electrons interact with the sample during a few femtoseconds, creating neutral electron-hole pairs. Because of the electrons high speed and the thin samples used (contrary to conventional CL schemes in Scanning Electron Microscope, SEM, on thick samples), the inelastic interaction is very small, leading to the creation of typically zero or one electron-hole pair above the diamond energy band gap per incident electron. The typical current in our experiments is of the order of 100 pA, or 1.6 electron every nanosecond, an order of magnitude lower than the NV$^{\text{0}}$ typical lifetimes. In this sense, the excitation of the NV$^{\text{0}}$ can be seen as continuous. The electron-hole pair rapidly (few picoseconds \cite{Yu2005}), loses energy to reach the bottom (top) of the conduction (valence) band. It may then diffuse up to an emission center, where it can recombine in the form of a photon. This creates a region around the center from which it can be excited, giving rise to maxima in light emission, as observed in previous experiments \cite{Tizei2012}. The described excitation scheme, together with the here proven SP detection mimicking closely PL experiments, sets CL in a STEM as the nanometre scale counterpart of PL. 

This unique capability to monitor the excitation position with subwavelength resolution on a nanostructure imaged in parallel is the key difference between the excitation of color centers using a fast electron beam and a laser (explained in Fig. \ref{figure3}d). The same advantage has been recently demonstrated in electron optical absorption experiments (electron energy-loss spectroscopy) \cite{deAbajo2010, Nelayah2007} or classical luminescence experiments in an electron microscope \cite{Zagonel2011}.

To evidence this advantage we have measured $g^{(2)}(\tau)$ at a second area (20 nm by 34 nm) on the very same diamond nanoparticle, separated by 130 nm from the first area (both marked on Fig. \ref{figure3}a-b). The separation between these positions is about $\lambda/5$ ($\lambda$ the wavelength of light). The values of $g^{(2)}(\tau = 0)$ are $(0.46 \pm 0.05)$ and $(0.73 \pm 0.02)$, respectively (Fig. \ref{figure3}c). The second value might be associated with a light emitted by two NV$^{\text{0}}$centers (or one NV$^{\text{0}}$ and some other potentially non SP emitter centers emitting at the same wavelength), as the antibunching dip should be shallower in these cases. This shows the remarkable ability to measure the temporal statistics properties of non-classical light beams excited from two distinct positions separated by subwavelength distances within the same nano-object. Moreover, this result proves that individual point defects can be detected with high spatial resolution using cathodoluminescence, as previously conjectured by us \cite{Tizei2012}.

The results presented here open the way to new research paths. From one perspective, we have demonstrated the ability to identify and count quantum emitters in close proximity and to measure their individual responses. This will allow a much better understanding of the physics of interaction between two or more emitters, should they be point defects, atoms or densely packed quantum dots \cite{Kako2006, Spinicelli2009}. For example, one may probe how the $g^{(2)}(\tau)$ function varies spatially as the excitation probe is scanned between two centers. This would render feasible new, otherwise impossible, quantum nano-optics experiments. In particular, the here proven spatial resolution is limited by the diffusion distance here, this is typically less than 100 nm \cite{Tizei2012}, but resolutions down to 5 nm are expected for other nanostructured systems \cite{Zagonel2011}. Such spatial resolution may allow the unique capability to image, to characterize and to address individual quantum bits in compact systems. In addition to pure imaging, electron microscopes now allow the study of chemical, electronic, magneto- and electrostatic properties of materials at the atomic scale, that could be used in parallel to $g^{(2)}(\tau)$ measurements. Naturally, such possibility would aid the characterization of future scalable quantum computing system.

Furthermore, our experiments reveal the emission of single photon fields excited by fast electrons. Therefore, our work represents a new approach to the generation of single photons states and could be applied to the study of electron-photon entanglement, as proposed recently \cite{Bendana2011}. Finally, the use of a pulsed electron source \cite{Bostanjoglo2000, Browning2012, Williamson1997} could lead to the creation of nanometre sized triggered SPS. We believe that this experimental approach may have a significant impact in SPS studies, just as electron energy-loss studies have greatly aided the comprehension of plasmons physics in metallic nano-objects.

Finally, fast electrons couple effectively to plasmons in nanometre wide dissipative metallic particles \cite{deAbajo2010, Nelayah2007}. Therefore, the study of the temporal statistics of light beams using fast electrons may allow new experiments in quantum plasmonics at the nanometre scale, in analogy to experiments performed using standard optical techniques \cite{Kolesov2009}. With the technique described here access to much smaller nanoparticles will allow the study of the quantum properties of plasmons in a highly dissipative regime.
We view this experiment as a shift toward deeply sub-wavelength quantum optics that will allow otherwise impossible precise characterization of quantum optical properties of confined objects and quantum emitter/plasmon coupling. The experiment presented can be implemented in other widely available TEM and STEM assuming a suitably designed light collection system \cite{Zagonel2011} is used.

We thank Marcel Tenc\'e for help in setting up the SPADS intensity mapping. We acknowledge Mike Walls, Andréia N. S. Hisi, Fran\c cois Treussart, Christian Colliex and Odile St\'ephan for critical reading of the manuscript. Luiz Zagonel for interesting discussions. This work has received support from the National Agency for Research under the program of future investment TEMPOS-CHROMATEM with the reference ANR-10-EQPX-50. The authors acknowledge financial support from the European Union under the Framework 7 program under a contract for an Integrated Infrastructure Initiative. The research leading to these results has received funding from the European Union Seventh Framework Programme [FP7/2007- 2013] under grant agreement n°312483 (ESTEEM2).


\end{document}